\documentstyle[twocolumn,aps]{revtex}

\begin{document}
\title{Thermal magnetization fluctuations in thin films and a new physical form for
magnetization damping}
\author{Vladimir L. Safonov and H. Neal Bertram}
\address{Center for Magnetic Recording Research, University of California - San\\
Diego, 9500 Gilman Drive, La Jolla, CA 92093-0401}
\date{\today}
\maketitle

\begin{abstract}
The effect of thermal fluctuations on a thin film magnetoresistive element
has been calculated. The technique involves adding to the basic spin
dynamics a general form of interaction with a thermal bath. For a general
anisotropic magnetic system the resulting equation can be written as a
Langevin equation for a harmonic oscillator. Our approach predicts two times
smaller noise power at low frequencies than the conventional stochastic
Landau-Lifshitz-Gilbert equation. It is shown that equivalent results can be
obtained by introducing a tensor phenomenological damping term to the
gyromagnetic dynamics driven by a thermal fluctuating field.
\end{abstract}


Recently Smith and Arnett \cite{smith}\ have reported that thermal
magnetization noise may be a serious problem for giant magnetoresistive
spin-valve heads designed for high areal storage density. Using the
phenomenological Landau-Lifshitz-Gilbert (LLG) equation with random fields
and the fluctuation-dissipation theorem, they obtained an estimate for the
low-frequency noise intensity. They considered the GMR head as a
single-domain anisotropic magnetic particle.

The goal of this Note is to show that the analysis in the framework of
conventional stochastic LLG equation gives an overstated noise level due to
inconsistency of isotropic phenomenological damping with basic dynamic
equations for anisotropic magnetic system. Using normal mode approach, we
derive a formula for the noise power spectrum, which at low frequencies is
two times smaller than theoretical estimate of Ref.\cite{smith}. We also
show that this result may be obtained in the framework of stochastic
gyromagnetic equation with tensor phenomenological damping.

Thermal noise is a phenomenon that is a result of microscopic spin
fluctuations due to interactions with a thermal bath. It has been customary
to include thermal fluctuations by simply adding random fields to dynamic
equations with phenomenological damping, such as the LLG equation \cite
{brown}. However the gyromagnetic term describes an averaged precession of a
large number of spins. A direct connection with microscopic physics must be
found for the phenomenological damping term.

Lax \cite{lax}\ has shown that independent of the specific coupling to the
heat bath, the resulting dynamic equation in the vicinity of equilibrium has
the form of a damped harmonic oscillator driven by a fluctuating force. Here
we show that for small fluctuations, an anisotropic magnetic dynamic system
can be transformed into the general form analyzed by Lax. We do this by a
series of coordinate transformations from the original magnetic variables to
a harmonic oscillator. By back transformation we have found a tensor form of
the phenomenological damping for the magnetization dynamic equation. With
this new form random fields may now be added to this dynamic equation, while
properly characterizing the physics of interaction with a thermal bath.

Consider thermal fluctuations in a single-domain magnetic particle with the
following energy density: 
\begin{equation}
{\cal E}/V_{0}=-K_{\parallel }(M_{z}/M_{{\rm s}})^{2}+K_{\perp }(M_{x}/M_{%
{\rm s}})^{2}-M_{z}H_{0}.  \label{energy}
\end{equation}
Here $\,V_{0}\,$ is the particle volume, $K_{\parallel }$ and $K_{\perp }$\
are the easy ($z$) and hard ($x$) axis anisotropy constants, respectively, $%
{\bf M=}(M_{x},M_{y},M_{z})$ is the magnetization $(|{\bf M}|=M_{{\rm s}})$
and $H_{0}$ is the external magnetic field oriented along the easy axis. For
a thin film $y$ and $z$ are in the plane.

For convenience of calculations we introduce the classical spin ${\bf S}=-%
{\bf M}V_{0}{\bf /}\hbar \gamma $, where $\gamma $ is the gyromagnetic
ratio. Planck's constant $\hbar $ is used as a dimensional parameter (it
could be another parameter of the same dimensionality, e.g., $M_{{\rm s}%
}V_{0}/\gamma $). All our analysis is pure classical and final result does
not contain $\hbar $. We shall describe small oscillations of the
magnetization in terms of complex variables$\;a^{\ast },\,a$, which are
classical analogs of creation and annihilation operators of a harmonic
oscillator. This technique is used for a description of spin waves (see,
e.g., \cite{lvov}). Complex variables can be introduced by an expanded
Holstein-Primakoff transformation \cite{hopri}: 
\begin{equation}
\frac{S_{x}}{\sqrt{S/2}}\simeq \frac{a-a^{\ast }}{i},~\frac{S_{y}}{\sqrt{S/2}%
}\simeq a+a^{\ast },~S_{z}=a^{\ast }a-S.  \label{h-pa}
\end{equation}
The quadratic part of the energy ({\ref{energy}}) can be written as

\begin{equation}
{\cal E}^{(2)}/\hbar =Aa^{\ast }a+(B/{2})(aa+a^{\ast }a^{\ast }),
\label{quadform}
\end{equation}
where $\,A=\gamma (H_{{\rm K}}^{(\parallel )}+H_{0}+H_{{\rm K}}^{(\perp
)}/2)\,$, $\,B=-\gamma H_{{\rm K}}^{(\perp )}/2$, $H_{{\rm K}}^{(\parallel
)}\equiv 2K_{\parallel }/M_{{\rm s}}$ and $H_{{\rm K}}^{(\perp )}\equiv
2K_{\perp }/M_{{\rm s}}$. To eliminate the nondiagonal terms from (\ref
{quadform}) we shall use the linear canonical transformation: 
\begin{eqnarray}
a=uc+vc^{\ast }, &\quad &a^{\ast }=uc^{\ast }+vc,  \label{canonic} \\
u=\sqrt{\frac{A+\omega _{0}}{2\omega _{0}}}, &\quad &v=-\frac{B}{|B|}\sqrt{%
\frac{A-\omega _{0}}{2\omega _{0}}},  \nonumber
\end{eqnarray}
where $\omega _{0}=\sqrt{A^{2}-B^{2}}$. Thus, we find the normal mode of the
system and the energy of an harmonic oscillator 
\begin{eqnarray}
{\cal E}^{(2)}/\hbar &=&\omega _{0}c^{\ast }c,  \label{ener} \\
\omega _{0} &=&\gamma \sqrt{(H_{{\rm K}}^{(\parallel )}+H_{0}+H_{{\rm K}%
}^{(\perp )})(H_{{\rm K}}^{(\parallel )}+H_{0})},  \label{omega}
\end{eqnarray}
where $\omega _{0}$\ is the ferromagnetic resonance frequency.

Using the energy (\ref{ener}), the dynamic equation for the complex
amplitude $c(t)$ is:

\begin{equation}
dc/dt=-i\omega _{0}c  \label{gyro}
\end{equation}
and is of a form that corresponds exactly to small amplitude gyromagnetic
rotation. Now we consider the interaction of this harmonic oscillator with a
reservoir (thermal bath)

\begin{equation}
{\cal E}^{(int)}/\hbar =c\sum_{k}g_{k}^{\ast }+c^{\ast }\sum_{k}g_{k},
\label{interaction}
\end{equation}
which will introduce two related effects: damping and fluctuations. Here $%
g_{k},\ g_{k}^{\ast }$ are the reservoir variables and incorporate specific
physical processes. Such interaction could have been added to the energy in
the $a,$ $a^{\ast }$ variables. However the linear transformation (\ref
{canonic}) does not alter this general form. Following \cite{lax},\cite
{scully}, the oscillator dynamics for this general form (\ref{interaction})
can be represented by the following Langevin equation:

\begin{equation}
dc/dt=-(i\omega _{0}+\eta )c+f(t),  \label{Langevin}
\end{equation}
where $\eta $ is defined by reservoir parameters and $f(t)$ is the random
force related to $\eta $. It should be noted that Eq.(\ref{Langevin}) is an
exact classical analog of a corresponding quantum mechanical equation \cite
{lax}. Note also that phenomenological introduction of dissipation by
assigning an imaginary part to the resonant frequency $\omega _{0}\pm i\eta $
is a common procedure in the theory of spin-wave instabilities \cite
{schloemann}.

We can start with Eq.(\ref{Langevin}) and rewrite it as a stochastic
differential equation with a damping parameter $\eta $ and a
thermodynamically consistent random field with diffusion coefficient $%
D=(\eta /\omega _{0})k_{{\rm B}}T/\hbar $ \cite{safonov},\cite{gardiner}:

\begin{equation}
dc=-(i\omega _{0}+\eta )cdt+\sqrt{D}[dW_{1}(t)+idW_{2}(t)].
\label{stochastic}
\end{equation}
Here $T$ is the temperature and $W_{1}(t)$ and $W_{2}(t)$ are two
independent Wiener processes. The solution of (\ref{stochastic})\ is

\begin{equation}
\frac{c(t)}{\sqrt{D}}=\int_{-\infty }^{t}e^{-(i\omega _{0}+\eta
)(t-t_{1})}[dW_{1}(t_{1})+idW_{2}(t_{1})].  \label{solution}
\end{equation}
Taking into account the rule of Ito's calculus (see, e.g., \cite{gardiner}) $%
dW_{m}(t_{1})dW_{n}(t_{2})=\delta _{mn}\Delta (t_{1}-t_{2})dt_{1}$, where $%
\Delta (t)$ is the Kronecker delta function, we can obtain the correlation
function with thermal averaging $\langle ...\rangle $:

\begin{equation}
\langle c^{\ast }(t)c(0)\rangle =(D/\eta )\exp (i\omega _{0}t-\eta t),
\label{correl}
\end{equation}
where $c^{\ast }(t)$ is the complex conjugate of $c(t)$.

The power spectral density of the complex amplitude fluctuations is defined
by

\begin{equation}
S_{c^{\ast }c}(\omega )=\int_{-\infty }^{\infty }\langle c^{\ast
}(t)c(0)\rangle e^{-i\omega t}dt.  \label{spectral}
\end{equation}
Substituting (\ref{correl}) into (\ref{spectral}) and including the correct
meaning of a damping process for both negative and positive time, one obtains

\begin{eqnarray}
S_{c^{\ast }c}(\omega ) &=&\frac{k_{{\rm B}}T}{\hbar \omega _{0}}%
\int_{-\infty }^{\infty }e^{i(\omega _{0}-\omega )t-\eta t\cdot {\rm sign}%
(t)}dt  \nonumber \\
&=&\frac{2(\eta /\omega _{0})k_{{\rm B}}T/\hbar }{(\omega _{0}-\omega
)^{2}+\eta ^{2}}.  \label{spectral1}
\end{eqnarray}

Let us consider now an application of our analysis to a magnetoresistive
head. We shall follow the model \cite{smith} of a GMR head as a thin film of
the volume $V_{0}$, uniformly magnetized along the $z$ axis of uniaxial
anisotropy. The transverse anisotropy is defined by shape demagnetization: $%
K_{\perp }\simeq 2\pi M_{{\rm s}}^{2}$. The power spectral density of random
fluctuations in the voltage $\delta V(t)$\ is defined by

\begin{equation}
S_{VV}(\omega )=\int_{-\infty }^{\infty }\langle \delta V(t)\delta
V(0)\rangle e^{i\omega t}dt.  \label{powerVV}
\end{equation}
The voltage deviation for a GMR head is $\delta V(t)=I_{{\rm bias}}\delta
R(t)$, which can be expressed in terms of the in-plane transverse
magnetization fluctuation

\begin{equation}
\delta V(t)=C_{V}\delta m_{x}(t),\quad C_{V}=I_{{\rm bias}}\frac{\partial R}{%
\partial H_{{\rm ext}}}\frac{\partial H_{{\rm ext}}}{\partial m_{x}}.
\label{volt}
\end{equation}
Thus, Eq.(\ref{powerVV}) can be rewritten as

\begin{equation}
S_{VV}(\omega )=C_{V}^{2}S_{m_{y}m_{y}}(\omega ),  \label{newpowerVV}
\end{equation}
where

\begin{equation}
S_{m_{y}m_{y}}(\omega )=\int_{-\infty }^{\infty }\langle \delta
m_{y}(t)\delta m_{y}\rangle (0)e^{i\omega t}dt  \label{mag-noise}
\end{equation}
defines the magnetization noise.

Since $\langle m_{y}(t)\rangle =0$, the deviation $\delta m_{y}(t)=m_{y}(t)$%
. Taking into account Eqs. (\ref{h-pa}) and (\ref{canonic}), one obtains

\begin{equation}
m_{y}=-S_{y}/S\simeq -(u+v)(c^{\ast }+c)/\sqrt{2S}.  \label{mx}
\end{equation}
Thus

\begin{eqnarray*}
S_{m_{y}m_{y}}(\omega ) &=&\frac{(u+v)^{2}}{2S}\big[\int_{-\infty }^{\infty
}\langle c(t)c^{\ast }(0)\rangle e^{i\omega t}dt \\
&&+\int_{-\infty }^{\infty }\langle c^{\ast }(t)c(0)\rangle e^{i\omega t}dt%
\big].
\end{eqnarray*}
Utilizing (\ref{spectral}) and (\ref{spectral1}), we have

\begin{eqnarray}
S_{m_{y}m_{y}}(\omega ) &=&\frac{\gamma k_{{\rm B}}T}{M_{{\rm s}}V_{0}}%
\left( \frac{H_{{\rm K}}+4\pi M_{{\rm s}}+H_{0}}{H_{{\rm K}}+H_{0}}\right)
^{1/2}\frac{\eta }{\omega _{0}}F(\omega ),  \label{final} \\
F(\omega ) &=&\frac{1}{(\omega _{0}-\omega )^{2}+\eta ^{2}}+\frac{1}{(\omega
_{0}+\omega )^{2}+\eta ^{2}}.  \nonumber
\end{eqnarray}
Examination of (\ref{final}) indicates that the damping parameter $\eta $
affects linearly the low frequency noise $(\omega \ll $ $\omega _{0})$ and
broadens the resonance peak. However the total power is independent of $\eta 
$.

In the case of small frequencies $\omega \ll $ $\omega _{0}\simeq \gamma 
\sqrt{4\pi M_{{\rm s}}H_{{\rm K}}}$ for $H_{0}=0$ and $4\pi M_{{\rm s}}\gg
H_{{\rm K}}$. From (\ref{final}) we obtain the noise power (as measured for
positive frequencies):

\begin{equation}
NP\simeq 2S_{m_{y}m_{y}}(0)\simeq \eta k_{{\rm B}}T/[\pi (\gamma H_{{\rm K}%
}M_{{\rm s}})^{2}V_{0}].  \label{ourresult}
\end{equation}
The corresponding noise power of \cite{smith} (in our notation) is

\begin{equation}
NP^{{\rm (SA)}}\simeq 4\alpha k_{{\rm B}}T/[\gamma (H_{{\rm K}})^{2}M_{{\rm s%
}}V_{0}],  \label{smithresult}
\end{equation}
where $\alpha $\ is the Gilbert damping parameter. To compare (\ref
{ourresult}) and (\ref{smithresult}) we need to find a relation between $%
\eta $ and $\alpha $. This can be done by comparing the dynamic equation for
the transverse magnetization given in Ref.\cite{smith} and the corresponding
transverse magnetization equation in our approach. From \cite{smith} we have

\begin{equation}
d^{2}m_{y}/dt^{2}+\alpha 4\pi \gamma M_{{\rm s}}dm_{y}/dt+\omega
_{0}^{2}m_{y}=h(t),  \label{dynmxsmith}
\end{equation}
where $h(t)$ is a random ``force''. Corresponding to (\ref{stochastic})
dynamic equation is:

\begin{equation}
d^{2}m_{y}/dt^{2}+2\eta dm_{y}/dt+\omega _{0}^{2}m_{y}=h(t).  \label{dynmx}
\end{equation}
From (\ref{dynmxsmith}) and (\ref{dynmx}), we obtain the following relation:

\begin{equation}
\alpha =\eta /2\pi \gamma M_{{\rm s}}.  \label{relation}
\end{equation}
Substituting (\ref{relation}) into (\ref{smithresult}), one can see that at
low frequencies our theory (\ref{ourresult}) predicts 2 times smaller (3 dB
smaller) noise power than the result of \cite{smith}: $NP/NP^{{\rm (SA)}%
}=1/2 $. It should be also noted that the conventional approach with
stochastic LLG equation gives different noise frequency dependence than our
result (\ref{final}).

Now we demonstrate the inconsistency of the conventional phenomenological
damping term with basic equations for a harmonic oscillator and show how
this isotropic damping can be replaced by a tensor damping term to obtain
the above results.

From the LLG equation

\begin{equation}
\frac{d{\bf M}}{dt}=-\gamma {\bf M}\times {\bf H}_{{\rm eff}}-\frac{\alpha
\gamma }{M_{{\rm s}}}{\bf M}\times ({\bf M}\times {\bf H}_{{\rm eff}})
\label{LLG}
\end{equation}
one can derive the following linearized equations:

\begin{eqnarray}
&&\frac{d}{dt}\left( 
\begin{array}{c}
m_{x} \\ 
m_{y}
\end{array}
\right)  \label{mxmyLLG} \\
&=&\left( 
\begin{array}{cc}
-\alpha \gamma (H_{{\rm K}}+4\pi M_{{\rm s}}+H_{0}) & -\gamma (H_{{\rm K}%
}+H_{0}) \\ 
\gamma (H_{{\rm K}}+4\pi M_{{\rm s}}+H_{0}) & -\alpha \gamma (H_{{\rm K}%
}+H_{0})
\end{array}
\right) \left( 
\begin{array}{c}
m_{x} \\ 
m_{y}
\end{array}
\right) ,  \nonumber
\end{eqnarray}
On the other hand, we can obtain the linearized equation for $m_{x}$ and $%
m_{y}$ from the normal mode equations (Eq.(\ref{stochastic}) without the
thermal term)

\begin{equation}
(d/dt+\eta )c=-i\omega _{0}c,\quad (d/dt+\eta )c^{\ast }=i\omega _{0}c^{\ast
}.  \label{normalmode}
\end{equation}
Substituting $c=ua-va^{\ast },~c^{\ast }=ua^{\ast }-va$ into (\ref
{normalmode}) and taking into account (\ref{h-pa}) and ${\bf m}=-{\bf S}/S$,
we obtain

\begin{eqnarray}
&&\frac{d}{dt}\left( 
\begin{array}{c}
m_{x} \\ 
m_{y}
\end{array}
\right)  \label{mxmy} \\
&=&\left( 
\begin{array}{cc}
-\eta & -\gamma (H_{{\rm K}}+H_{0}) \\ 
\gamma (H_{{\rm K}}+4\pi M_{{\rm s}}+H_{0}) & -\eta
\end{array}
\right) \left( 
\begin{array}{c}
m_{x} \\ 
m_{y}
\end{array}
\right) .  \nonumber
\end{eqnarray}
By comparison with (\ref{mxmyLLG}) we see that the nondiagonal terms
coincide. The diagonal terms in (\ref{mxmyLLG}) and (\ref{mxmy}),
responsible for relaxation, differ from each other.

In order to have the magnetization equations (\ref{mxmy}), one should start
with the magnetization equation containing a tensor damping of the form:

\begin{eqnarray}
\frac{d{\bf M}}{dt} &=&-\gamma {\bf M}\times {\bf H}_{{\rm eff}}-\gamma 
\frac{{\bf M}}{M_{{\rm s}}}\times \lbrack \stackrel{\leftrightarrow }{\alpha 
}\cdot ({\bf M}\times {\bf H}_{{\rm eff}})],  \label{LL} \\
\stackrel{\leftrightarrow }{\alpha } &=&\frac{\eta \gamma }{\omega _{0}^{2}}%
\left( 
\begin{array}{ccc}
H_{{\rm K}}^{(\parallel )}+H_{0}+H_{{\rm K}}^{(\perp )} & 0 & 0 \\ 
0 & H_{{\rm K}}^{(\parallel )}+H_{0} & 0 \\ 
0 & 0 & H_{0}
\end{array}
\right) .  \label{alpha}
\end{eqnarray}
For small oscillations about equilibrium, the $\alpha _{zz}$ component does
not enter into the equations. The conventional LLG equation can be used only
for a uniaxial system, where $H_{{\rm K}}^{(\perp )}=0$ and the tensor is
effectively a scalar.

We thank H. Suhl, C. E. Patton and N. Smith for valuable discussions, and X.
Wang for his help with the analysis of paper \cite{smith}. This work was
partly supported by matching funds from the Center for Magnetic Recording
Research at the University of California - San Diego and CMRR incorporated
sponsor accounts.

\end{document}